\documentclass{vgtc}                          % final (conference style)
\graphicspath{{figures/}{pictures/}{images/}{./}} % where to search for the images

\usepackage{times}                     % we use Times as the main font
\usepackage{tabu}                      % only used for the table example
\usepackage{booktabs}                  % only used for the table example
\usepackage{lipsum}                    % used to generate placeholder text
\usepackage{mwe}                       % used to generate placeholder figures

\usepackage{mathptmx}                  % use matching math font

\usepackage{amsmath}

\usepackage{multirow}

\usepackage{makecell}

\usepackage{enumitem}
\setitemize[1]{itemsep=0pt, partopsep=0pt, parsep=\parskip, topsep=5pt}

\newcommand{\RV}[2]{{#2}}

\newcommand{\change}[1]{#1}

\newcommand{\reducesectionspace}{\vspace{-3pt}}
\newcommand{\reducesubsectionspace}{\vspace{-3pt}}

\usepackage{float}

\usepackage[toc]{appendix}

\usepackage{tablefootnote}

\onlineid{1619}

\vgtccategory{Research}

\vgtcinsertpkg
\crefformat{footnote}{#2\footnotemark[#1]#3}

\title{\change{XR Reality Check: What Commercial Devices Deliver for Spatial Tracking}}
\author{Tianyi Hu\\ \scriptsize \makecell{Department of Electrical and\\ Computer Engineering} \\\scriptsize Duke University %
\and Tianyuan Du\\ \scriptsize \makecell{Department of Electrical and\\ Computer Engineering} \\\scriptsize Duke University %
\and Zhehan Qu\\ %
     \scriptsize Department of Computer Science \\\scriptsize Duke University %
\and Maria Gorlatova\thanks{e-mail: \{tianyi.hu,alex.du,zhehan.qu,maria.gorlatova\}@duke.edu} \\ \scriptsize \makecell{Department of Electrical and\\ Computer Engineering} \\\scriptsize Duke University}

\teaser{
  \includegraphics[width=1.03\linewidth]{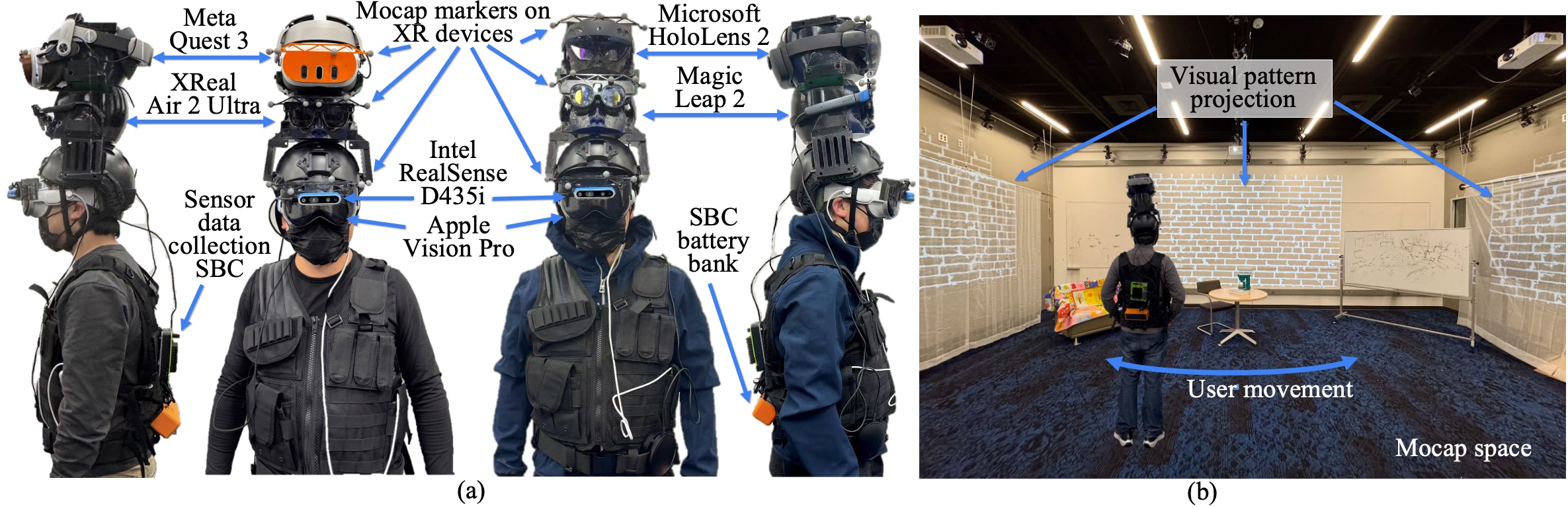}
  \vspace{-0.9cm}
  \caption{(a) XR devices and data collection apparatus, and (b) data collection environment.}
  \label{fig:teaser}
  \vspace{-0.3cm}
}

\abstract{
\begin{abstract} 

Inaccurate spatial tracking in extended reality (XR) devices leads to virtual object jitter, misalignment, and user discomfort, fundamentally limiting immersive experiences and natural interactions. In this work, we introduce a novel testbed that enables simultaneous, synchronized evaluation of multiple XR devices under identical environmental and kinematic conditions. Leveraging this platform, we present the first comprehensive empirical benchmarking of five state-of-the-art XR devices across 16 diverse scenarios. Our results reveal substantial intra-device performance variation, with individual devices exhibiting up to 101\% increases in error when operating in featureless environments. We also demonstrate that tracking accuracy strongly correlates with visual conditions and motion dynamics. We also observe significant inter-device disparities, with performance differences of up to 2.8$\times$, which are closely linked to hardware specifications such as sensor configurations and dedicated processing units. Finally, we explore the feasibility of substituting a motion capture system with the Apple Vision Pro as a practical ground truth reference. While the Apple Vision Pro delivers highly accurate relative pose error estimates ($R^2 = 0.830$), its absolute pose error estimation remains limited ($R^2 = 0.387$), highlighting both its potential and its constraints for rigorous XR evaluation. This work establishes the first standardized framework for comparative XR tracking evaluation, providing the research community with reproducible methodologies, comprehensive benchmark datasets, and open-source tools that enable systematic analysis of tracking performance across devices and conditions, thereby accelerating the development of more robust spatial sensing technologies for XR systems.
\end{abstract}
} % end of abstract

\keywords{Human-centered computing—Ubiquitous and mobile computing design and evaluation methods;  Mixed/augmented reality; Computing methodologies—Computer vision problems; Tracking; Empirical studies}

\begin{document}


%%\firstsection{Introduction}

\maketitle

\section{Introduction}
\label{sec:Introduction}
\reducesectionspace
The rapid advancement of Extended Reality (XR) technologies has generated significant interest across research, development, and consumer domains. Contemporary XR systems predominantly leverage inside-out tracking methodologies that employ onboard camera arrays and inertial measurement units (IMU) to simultaneously estimate user motion and reconstruct environmental geometry ~\cite{klein2007parallel,qin2018vins,klein2009parallel,metaSLAM,li2024rd}. This approach allows on-device tracking of user pose, which offers improved portability and usability compared to traditional outside-in tracking systems that require fixed external infrastructure~\cite{jinyu2019survey,OculusRift,SteamVR}. However, inherent limitations persist in visual-inertial odometry (VIO) and visual-inertial SLAM (VI-SLAM) implementations, particularly under challenging operational conditions including high rotational velocities, low-light environments, and textureless spaces. 
% These factors can induce tracking errors that manifest as holographic drift and positional instability, ultimately degrading immersive experiences~\cite{OculusRift,OculusQuest,scargill2021will}. 
These factors can induce tracking errors that manifest as drift and misalignment of virtual content, resulting in positional instability that ultimately degrades the immersive experience~\cite{OculusRift,OculusQuest,scargill2021will}. 
While recent advancements have improved tracking robustness~\cite{jinyu2019survey,sheng2024review,rabbi2013survey}, SOTA XR devices continue to demonstrate performance limitations, as evidenced by empirical user reports and developer documentation~\cite{TheWiredReview,TheDigitalTrendsReview,lumafield}.

A rigorous quantitative evaluation of XR tracking systems is critical for developers optimizing immersive applications and users selecting devices. However, three fundamental challenges impede systematic performance analysis across commercial XR platforms.
%%%%%%%%%%%%%%%%%%%%%%%%%%%%%%%%%%%%%%%%
Firstly, major XR manufacturers do not reveal critical tracking performance metrics, sensor (tracking camera and IMU) interfaces, or algorithm architectures. 
This lack of transparency prevents independent validation of tracking reliability and limits decision-making by developers and end users alike.
%%%%%%%%%%%%%%%%%%%%%%%%%%%%%%%%%%%%%%%%
Secondly, existing studies~\cite{OculusRift, OculusQuest, AWE2021, ViconQuest2, SteamVR, hu2024apple} have either focused on single-device analysis or evaluated multiple devices individually without standardized methodologies for cross-device comparisons under controlled environmental and kinematic conditions.
%%%%%%%%%%%%%%%%%%%%%%%%%%%%%%%%%%%%%%%%
Thirdly, existing evaluations focus on trajectory-level performance but omit correlation analyses at timestamp level that link pose errors to camera and IMU sensor data. This omission limits the ability to analyze how environmental factors and user kinematics influence estimation accuracy.
%%%%%%%%%%%%%%%%%%%%%%%%%%%%%%%%%%%%%%%%
\textcolor{black}{Finally, most prior work does not share testbed designs or experimental datasets, limiting reproducibility, validation, and subsequent research, such as efforts to model, predict, or adapt to pose errors based on trajectory and sensor data.}

In this work, we propose a novel XR spatial tracking testbed that addresses all the aforementioned challenges. The testbed enables the following functionalities: (1) synchronized multi-device tracking performance evaluation under various motion patterns and configurable environmental conditions; (2) quantitative analysis among environmental characteristics, user motion dynamics, multi-modal sensor data, and pose errors; and (3) open-source calibration procedures, data collection frameworks, and analytical pipelines.

We validate our testbed through comprehensive experiments with five commercial XR devices, 
including Apple Vision Pro~\cite{Apple}, Meta Quest 3~\cite{Meta}, HoloLens 2~\cite{HoloLens2}, Magic Leap 2~\cite{MagicLeap}, and XReal Air 2 Ultra~\cite{XReal},  across diverse operational scenarios.  
\change{
Furthermore, our analysis reveal that the Apple Vision Pro's  tracking accuracy (with an average relative pose error (RPE) of 0.52~cm, which is the best among all) enables its use as a ground truth reference for evaluating other devices' RPE without the use of a motion capture system.
}
%%%%%%%%%%
\RV{R1C4, R2C1}{The full implementation, including hardware designs, software code, benchmarking tools, \textcolor{black}{and experimental dataset,} is publicly accessible via an open-source repository\footnote{\label{GitHubRepo}\href{https://github.com/Duke-I3T-Lab/XR_Tracking_Evaluation}{github.com/Duke-I3T-Lab/XR\_Tracking\_Evaluation}} 
to promote reproducibility and standardized evaluation in the XR research community.}
Our main contributions are as follows:
%\begin{myitemize}
\begin{itemize}[itemsep=0pt, topsep=0.3pt, parsep=1pt, leftmargin=0.15in]
    \item Designed a novel testbed enabling simultaneous evaluation of multiple XR devices under the same environmental and kinematic conditions. This testbed achieves accurate evaluation via time synchronization precision and extrinsic calibration.    
    \item 
    \change{Conducted the first comparative analysis of five SOTA commercial XR devices (four headsets and one pair of glasses), quantifying spatial tracking performance across 16 diverse scenarios. Our analysis reveals that average tracking errors vary by up to 2.8× between devices under identical challenging conditions, with errors ranging from sub-centimeter to over 10~cm depending on devices, motion types, and environment conditions.} 
    
    \item 
    \change{Performed correlation analysis on collected sensor data to quantify the impact of environmental visual features, SLAM internal status, and IMU measurements on pose error, demonstrating that different XR devices exhibit distinct sensitivities to these factors.}

    \item 
    \change{Presented a case study evaluating the feasibility of using Apple Vision Pro as a substitute for traditional motion capture systems in tracking evaluation. Our results show that relative pose error measurements from Apple Vision Pro strongly correlate with those from the motion capture system ($R^2 = 0.830$). However, while the correlation for absolute pose error is substantially lower ($R^2 = 0.387$), this suggests that Apple Vision Pro provides a reliable reference for local tracking accuracy, making it a practical tool for many XR evaluation scenarios despite its limitations in assessing global pose precision.}
     
%\end{myitemize}
\end{itemize}

\change{
The remainder of this paper is organized as follows: Section~\ref{sec:RelatedWork} reviews related work. Section~\ref{sec:SystemDesign} details the design of our testbed system. Section~\ref{sec:ExperimentSetup} describes the experimental setup, and Section~\ref{sec:ExperimentResult} presents the experimental results. Section~\ref{sec:CaseStudy} provides a case study demonstrating the potential of using the Apple Vision Pro as a ground truth reference in place of a mocap system. Section~\ref{sec:DiscussionImplication} discusses the implications of our findings, while Section~\ref{sec:FutureWork} addresses the limitations of this work and suggests directions for future research. Finally, Section~\ref{sec:Conclusions} concludes the paper.
}

\reducesectionspace

\section{Related Work}
\label{sec:RelatedWork}
\reducesectionspace
\textbf{Factors influencing XR tracking performance}:
Modern visual-inertial odometry (VIO) and visual-inertial simultaneous localization and mapping (VI-SLAM) systems face notable challenges in tracking accuracy, which are predominantly shaped by environmental and kinematics conditions. Texture-deficient surfaces impair environmental perception, 
limiting the ability of VI-SLAM to extract meaningful features for localization~\cite{bujanca2021robust, jinyu2019survey,macario2022comprehensive,garforth2019visual,guo2012analysis}. 
In contrast, environments with high structural complexity and distinct visual features significantly improve pose estimation accuracy~\cite{schonberger2016structure, ferranti2021can,zhang2021perceived,liu2019edge,rosenholtz2007measuring,gabbard2007active}.
Kinematic factors, such as abrupt rotations or translations of the headset, exacerbate performance degradation by inducing motion blur in visual data streams, which leads to feature point tracking loss~\cite{liu2021mba, nardi2015introducing,zhang2018tutorial,campos2021orb,rabbi2013survey}. To address these limitations, our evaluation framework systematically incorporates operational constraints to quantitatively assess XR device tracking performance under diverse controlled conditions.

\textbf{Existing XR tracking evaluation methods}:
Reliable spatial tracking is foundational for most modern XR systems, and extensive research exists on evaluation methodologies of spatial tracking. Among them, three approaches are widely adopted.

\textit{Robotic actuation platforms}: Robotic arms execute predefined kinematic profiles with high repeatability~\cite{OculusQuest, OculusRift,AWE2021,mulvany2020target}. While ensuring standardized testing, these methods impose artificial motion constraints and limited kinematic ranges that do not replicate natural human head movements, reducing ecological validity.

\textit{Holographic drift analysis}: Studies such as \cite{vassallo2017hologram, slocum2021realitycheck, scargill2022here,scargill2023ambient} measure virtual object displacement as a proxy for tracking error caused by pose estimation inaccuracies. However, this method is not a direct evaluation of the tracking errors and restricts evaluations to specific environments, limiting cross-device comparisons under diverse conditions.

\textit{Motion capture ground truthing}: Infrared marker-based systems provide millimeter-accurate reference trajectories (e.g., \cite{Vicon,ViconQuest2, SteamVR, monica2022evaluation,grupp2017evo}), making them a common choice for validating XR tracking systems. However, existing implementations often impose artificial constraints that limit their ecological validity. For example, Holzwarth et al.~\cite{SteamVR} restrict evaluations to 2D planar motion using trolley-mounted headsets, while Boulo et al.~\cite{ViconQuest2} focus on linear translation patterns. Monica et al.~\cite{monica2022evaluation} maintain static environmental parameters, not explicitly addressing dynamic interactions between user kinematics and environmental factors. Similarly, Hu et al.~\cite{hu2024apple} analyze devices individually without standardized methodologies for cross-device comparisons, limiting generalizability.

To the best of our knowledge, this study introduces the first XR testbed capable of systematically assessing tracking performance across multiple XR devices. The platform considers environmental factors, user kinematics, and sensor characteristics, evaluating performance at trajectory and timestamp levels. Additionally, we present results for five commercial XR devices: Apple Vision Pro, Meta Quest 3, Magic Leap 2, HoloLens 2, and XReal Air 2 Ultra. 

\reducesectionspace

\section{System Design}
\label{sec:SystemDesign}
\reducesectionspace
\begin{figure}[t]
\centering
\includegraphics[width=0.85\linewidth]{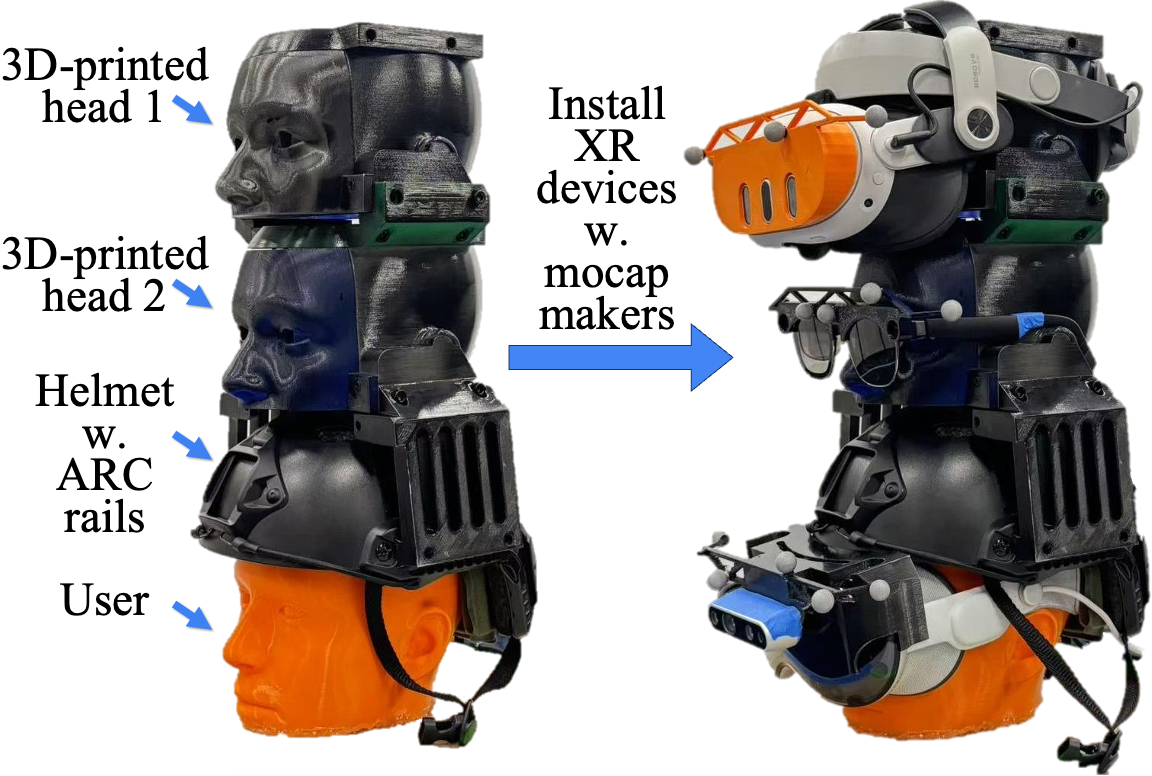}
\vspace{-0.5cm}
\caption{Setup of 3D-printed heads mounted on a helmet for simultaneous XR device evaluation.}
\vspace{-0.5cm}
\label{fig:3D-print-heads-on-helmet}
\end{figure}

\begin{figure*}[th!]
\centering
\includegraphics[width=0.97\linewidth]{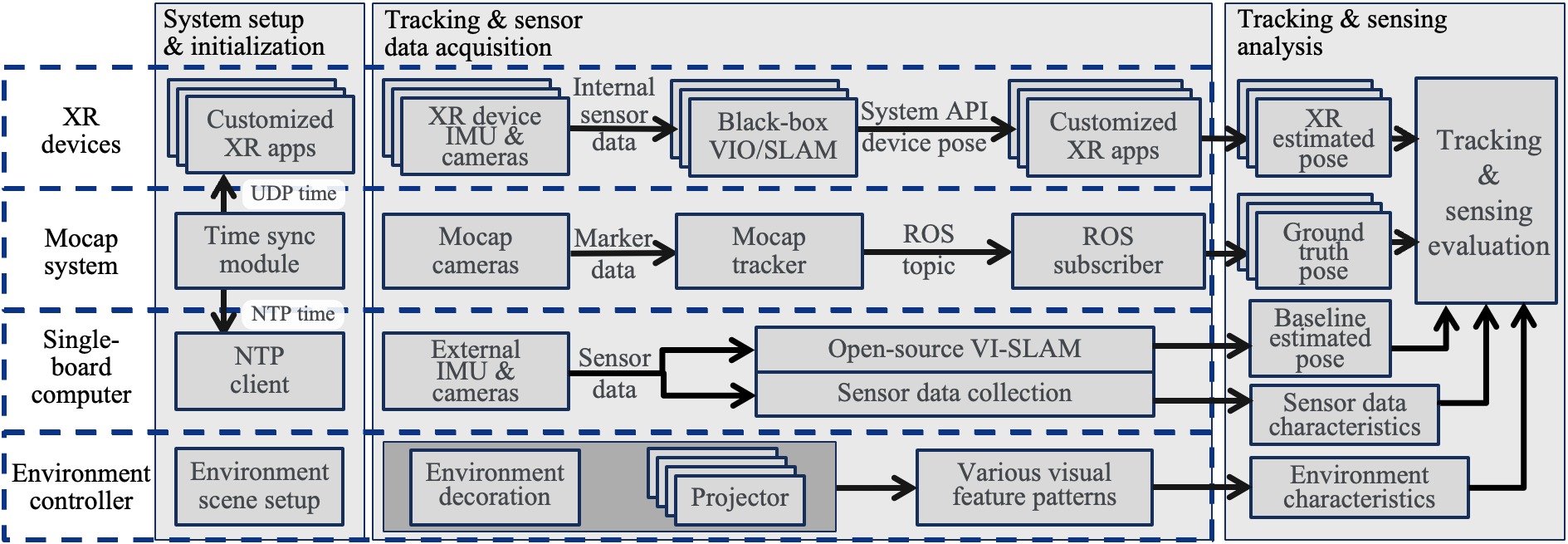}
\vspace{-0.4cm}
\caption{Testbed pipeline for multiple XR devices' tracking performance evaluation.}
\label{fig:TestbedSystemPipeline}
\vspace{-0.6cm}
\end{figure*}
 
The testbed system comprises four primary hardware components: (1) a set of target XR devices for evaluation, (2) a motion capture (mocap) station for collecting ground truth trajectories and ensuring time synchronization, and (3) a single-board computer with an Intel RealSense camera for sensor data collection and execution of an open-source VI-SLAM algorithm as a baseline.
\cref{fig:TestbedSystemPipeline} illustrates the testbed pipeline, structured into three main stages: System Setup and Initialization (\S~\ref{subsec:SystemSetupInit}), Tracking \& Sensor Data Acquisition (\S~\ref{subsec:TrackingSensingDataAcquisition}), and Tracking \& Sensing Analysis (\S~\ref{subsec:TrackingDataAnalysis}).

\reducesubsectionspace
\subsection{System Setup and Initialization}  
\label{subsec:SystemSetupInit}  
\reducesubsectionspace

The system setup and initialization phase integrates both hardware and software components to facilitate a comprehensive evaluation of XR tracking performance. The setup process is conducted once during the construction of the testbed or when introducing a new XR device for evaluation, whereas initialization is performed prior to each data collection round. This design enables simultaneous evaluation of multiple XR devices and sensor data acquisition. We develop a custom 3D-printed head structure, mounted on a helmet, to securely hold XR devices alongside the sensor data collection unit. 
To capture accurate ground-truth pose, infrared markers were affixed to the XR devices. Meanwhile, custom applications collected pose estimate output by each device. To ensure precise evaluations, extrinsic calibration of all XR devices was completed during setup, and time synchronization was performed during initialization before each data collection session. The following paragraphs provide detailed descriptions of these components.

\reducesubsectionspace
\subsubsection{Multi-XR Device Evaluation Platform Setup}
\reducesubsectionspace
Existing studies \cite{hu2024apple, OculusQuest, OculusRift, AWE2021}  evaluate XR devices individually, which introduces inconsistencies due to variations in user movement and environmental conditions across trials. To address this limitation, we developed a multi-device evaluation platform featuring two 3D-printed head structures mounted on a helmet equipped with accessory rail connectors (ARC). The 3D-printed platform was designed to vertically stack the 3D-printed head structures, ensuring that the field of view (FoV) of each XR device’s camera sensors remains unobstructed, thereby avoiding interference with tracking performance. This setup enables simultaneous evaluation of three XR devices under identical user movements and environmental conditions, ensuring fair comparisons.
%%%%%%%%%%%%%%
\cref{fig:3D-print-heads-on-helmet} and \cref{fig:teaser}(a) illustrate the platform, where two XR devices are mounted on the 3D-printed heads, while the third XR device is worn by the user. 

\begin{figure}[t]
\centering
\includegraphics[width=0.9\linewidth]{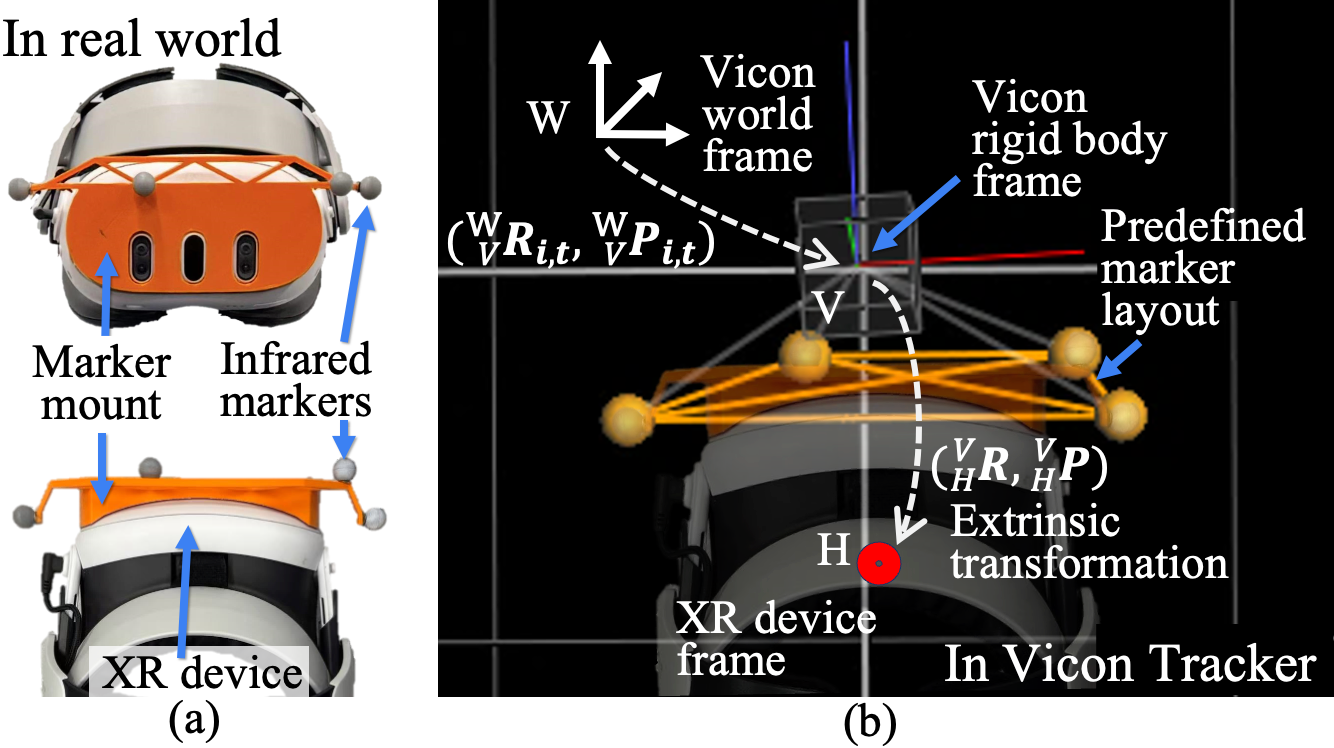}
\vspace{-0.4cm}
\caption{\RV{R3C7}{Infrared makers on the XR device and corresponding rigid body registration in the Vicon Tracker.}}
\vspace{-0.4cm}
\label{fig:MarkersAndTransformation}
\end{figure}

\reducesubsectionspace
\subsubsection{XR Devices Setup and Initialization}
\reducesubsectionspace

\textbf{Infrared markers \& extrinsic calibration:} 
\RV{R3C7}{
To capture precise ground truth trajectories for XR devices, we affix infrared markers to each device (\cref{fig:MarkersAndTransformation}), registering them as rigid bodies with a reference frame $V_{i,t}$ in the Vicon Tracker system for device $i$ at time $t$. Each device was assigned a unique marker layout to enable simultaneous multi-device tracking.
%%%%%%%
While this setup provided real-time 6-DoF ground truth pose data, accurate evaluation required aligning the reference frame $V_{i,t}$ in the Vicon with the XR device’s reference frame $H_{i,t}$, representing estimated orientation and translation of device $i$ at time $t$. As shown in \cref{fig:MarkersAndTransformation}(b), any misalignment between $V_{i,t}$ and $H_{i,t}$ introduces systematic errors in trajectory comparisons, potentially compromising evaluation accuracy.
}

\RV{R4C1\&R3C1}{However, XR device manufacturers do not specify the precise location of the device reference frame, making direct measurement of the offset infeasible. Additionally, these proprietary devices restrict developer access to sensor data, preventing the use of calibration toolboxes (e.g., Kalibr~\cite{rehder2016extending}) to compute extrinsic parameters from sensor readings.}

\RV{R4C1\&R3C1, R3C7}{
To resolve this, we compute a rigid transformation (${}^{V}_{H}\mathbf{P}_i \in \mathrm{SE}(3)$) that maps the Vicon rigid body frame $\mathbf{V}_{i,t}$ to the XR device frame $\mathbf{H}_{i,t}$ for each device $i$. This extrinsic calibration is performed once per device by recording synchronized pose sequences $\left( \mathbf{V}_{i,t}, \mathbf{H}_{i,t} \right)_{t = t_0}^{t_T}$ along a trajectory starting at time $t_0$ and ending at time $t_T$. Following established methods~\cite{OculusQuest, monica2022evaluation, grupp2017evo}, we iteratively optimize the transformation matrix ${}^{V}_{H}\mathbf{P}_i$ to minimize the discrepancy between the transformed ground truth trajectory, $\left( {}^{W}_{V}\mathbf{P}_{i,t} \cdot {}^{V}_{H}\mathbf{P}_i \right)_{t = t_0}^{t_T}$, and the estimated trajectory from the XR device, $\left( {}^{W}_{H}\mathbf{P}_{i,t} \right)_{t = t_0}^{t_T}$. Calibration is conducted in a feature-rich environment with slow, smooth head movements to minimize XR pose estimation errors, ensuring that residual discrepancies predominantly reflect transformation inaccuracies rather than tracking errors.}
    
After completing this one-time extrinsic calibration, we apply the transformation matrix to the ground truth trajectories obtained from the motion capture system. This yields ground truth trajectories that are accurately aligned with the device centers, thereby enhancing the reliability of tracking performance evaluations.

\textbf{XR time synchronization:} Accurate temporal alignment is essential for precise pose error evaluation in our XR tracking system, which involves multiple devices with independent internal clocks. Discrepancies in system time can lead to misaligned data and unreliable results. To address this challenge, we implemented tailored synchronization methods during the system initialization phase, ensuring consistent timing across all devices.
%%%%%%%%%%%%%%%%%%%
For XR devices, direct modification of device clocks was not possible due to restricted system privileges in our customized applications. We instead implemented a timestamp offset protocol using a server-client architecture over single-hop UDP. We measured the round-trip time of sending packets over UDP, which revealed an average latency of 10.42 ms, resulting in a one-way delay of 5.21 ms, which is negligible for trajectory evaluation. 
%%%%%%%%%%%%%%%%%%%
During initialization, each device executed a custom synchronization client that shared its IP address with the motion capture server. The server responded with a 10-second timestamp stream, enabling devices to compute local clock offsets relative to the reference time. These offsets were queued and averaged to mitigate network jitter, yielding a stable time delta applied throughout data collection. 
This approach ensured consistent temporal alignment between XR devices and the mocap system without requiring direct modification of the XR device clocks.

\begin{figure}[t!]
\centering
\includegraphics[width=0.92\linewidth]{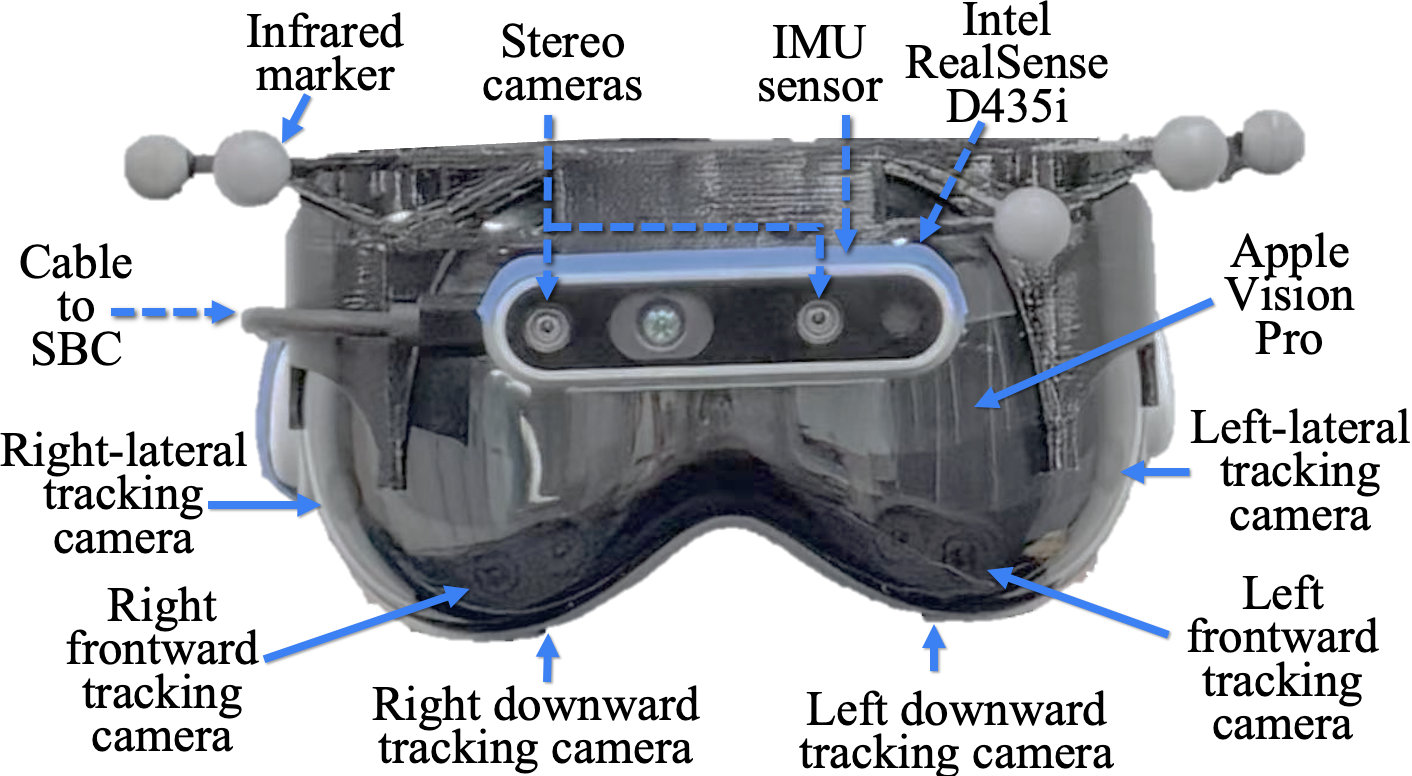}
\vspace{-0.4cm}
\caption{Apple Vision Pro with the sensor data collection module.}
\vspace{-0.6cm}
\label{fig:AVPandSensor}
\end{figure}

\reducesubsectionspace
\subsubsection{Sensor Data Collection Module Setup and Initialization}
\label{sec:SensorDataCollection}
\reducesubsectionspace

\textbf{Hardware setup:} 
\RV{R4C2\&R3C2, R3C5, R3C8, R1C3}{
To perform correlation analysis of the XR tracking performance with sensor characteristics and enable comparison with open-source VI-SLAM algorithms, we collected camera and IMU data streams during data acquisition. As most XR devices restrict developer access to their sensors, we employed the Intel RealSense D435i, which integrates dual monochrome global shutter cameras and an IMU, for capturing visual-inertial data and running open-source stereo-inertial SLAM. 
The RealSense was mounted on the AVP as illustrated in \cref{fig:AVPandSensor}, allowing the RealSense to leverage the AVP’s infrared markers for ground truth trajectory acquisition via the Vicon Tracker. We performed extrinsic calibration to compute the local transformation between the rigid body center and the sensor module center.
\textit{Importantly, the RealSense does not replace the AVP proprietary tracking module or obstruct the AVP's tracking cameras.} 
} During experiments, the sensor module was connected to an ODROID H2+ single-board computer (SBC) mounted on a tactical vest worn by the user (\cref{fig:teaser}), with a battery bank providing power to the SBC throughout the experiments.

\textbf{SBC time synchronization:} Accurate time synchronization between the SBC and the mocap server is essential to correctly interpret sensor readings in relation to tracking performance. Unlike XR devices, time synchronization on the SBC is straightforward due to administrative privileges provided by Ubuntu OS. We employed the Network Time Protocol (NTP) over a single-hop wireless local area network, with the mocap station serving as the NTP server. NTP achieves millisecond-level accuracy by calculating time offsets and round-trip delays between the client and server, providing high precision for the time synchronization.

\begin{table*}[t!]
\centering
\caption{XR device SDK for collecting the estimated pose.}
\vspace{-6 px}
\label{tab: XR-SDK-API-for-estimated-pose}
\begin{minipage}{\textwidth} \centering
\begin{tabular}{lll}
\textbf{XR Device} &
  \textbf{SDK} &
  \textbf{Pose anchor API} \\ \hline
Apple Vision Pro &
  \begin{tabular}[c]{@{}l@{}}VisionOS\\ on XCode\end{tabular} &
  \begin{tabular}[c]{@{}l@{}}deviceAnchor = worldTracking.queryDeviceAnchor(atTimestamp: CACurrentMediaTime()) \\ transformMatrix = deviceAnchor.originFromAnchorTransform\\ devicePosition= transformMatrix.columns.3 \\ deviceRotation = simd\_quatf(transformMatrix.upperLeft3x3)\end{tabular} \\ \hline
HoloLens 2 &
  \begin{tabular}[c]{@{}l@{}}Win10 SDK\\ on Unity\end{tabular} &
  \begin{tabular}[c]{@{}l@{}}Vector3 devicePosition = Camera.main.transform.position;\\ Quaternion deviceRotation = Camera.main.transform.rotation;\end{tabular} \\ \hline
Magic Leap 2 &
  \begin{tabular}[c]{@{}l@{}}OpenXR\\ on Unity\end{tabular} &
  \begin{tabular}[c]{@{}l@{}}headPositionAction = new InputAction(binding: ”/devicePosition”);\\ headRotationAction = new InputAction(binding: ”/deviceRotation”);\\ var devicePosition = headPositionAction.ReadValue\textless{}Vector3\textgreater{}();\\ var deviceRotation = headRotationAction.ReadValue\textless{}Quaternion\textgreater{}();\end{tabular} \\ \hline
Meta Quest 3 &
  \begin{tabular}[c]{@{}l@{}}Oculus XR\\ Plugin\\ on Unity\end{tabular} &
  \begin{tabular}[c]{@{}l@{}}OVRCameraRig cameraRig; \footnote{\RV{R1C1}{We use the center eye anchor located in between the left and right eye anchor as the Meta Quest 3's device anchor.}}
  \\ Vector3 devicePosition = cameraRig.centerEyeAnchor.position;\\ Quaternion deviceRotation = cameraRig.centerEyeAnchor.rotation;\end{tabular} \\ \hline
XReal Air 2 Ultra &
  \begin{tabular}[c]{@{}l@{}}NRSDK\\ on Unity\end{tabular} &
  \begin{tabular}[c]{@{}l@{}}devicePose = NRFrame.HeadPose;\\ var devicePosition = devicePose.position;\\ var deviceRotation = devicePose.rotation;\end{tabular} \\ \hline
\end{tabular}
\end{minipage}
\vspace{-0.4cm}
\end{table*}

\reducesubsectionspace
\subsection{Tracking and Sensor Data Acquisition}  
\label{subsec:TrackingSensingDataAcquisition}
\reducesubsectionspace
During the tracking and sensor data acquisition phase, six trajectories are simultaneously recorded: three ground truth trajectories from the mocap system for XR devices and three estimated trajectories from the devices themselves. Additionally, sensor data streams are collected using an Intel RealSense camera connected to a single-board computer attached to the user.

\textbf{Mocap system:}  
The mocap system tracks infrared markers attached to XR devices, associating them with predefined rigid body layouts established during setup. 
\RV{R3C7}{
For each XR device~$i$, the system records orientation ${ }^{W}_{V}\mathbf{R}_{i,t}$ and translation ${ }^{W}_{V}\mathbf{P}_{i,t}$ relative to the world coordinate system at time $t$, sampled at 100 Hz. 
}
Data are stored in a CSV file with eight columns: timestamp, translations along X, Y, and Z axes, and quaternions representing rotation.

\textbf{XR devices:}  
XR devices estimate user motion using onboard tracking cameras and IMU sensors. Custom applications were developed to invoke device-specific APIs for trajectory estimation. For example, Apple Vision Pro uses ARKit’s queryDeviceAnchor(), while Meta Quest 3 utilizes Unity Engine’s ovrcamerarig.centerEyeAnchor. The SDK and API we used for collecting the estimated pose from all the XR devices we evaluated are listed in \cref{tab: XR-SDK-API-for-estimated-pose}. 
These APIs are queried at the screen refresh rate, and timestamps are adjusted using synchronization offsets determined during initialization. The estimated poses are saved in CSV files with a format consistent with that of the mocap system. Following data collection, these files are uploaded to the mocap server for centralized evaluation.
\RV{R2C1}{
Our XR data collection pipeline is implemented using Unity, OpenXR, and Xcode, which are widely supported across XR platforms. This extensible architecture facilitates easy adaptation to additional XR devices by updating the pose query API according to the relevant device SDK.
}

\textbf{Sensor data collection module:}
The sensor data collection module employs Intel RealSense D435i to capture synchronized data from multiple sensors, including an IMU and two infrared cameras which forms stereo cameras. The IMU consists of a gyroscope and an accelerometer, configured to sample data at 200 Hz. The stereo infrared cameras capture grayscale images at a resolution of 640 × 480 pixels and a frame rate of 30 FPS with auto-exposure enabled and the emitter disabled.
During the sensor data acquisition phase, the IMU data (linear acceleration and angular velocity) is logged in a CSV file with timestamps when both gyroscope and accelerometer streams are updated. Infrared frames are saved as PNG images with timestamps, accompanied by metadata in CSV files for subsequent evaluation.

\reducesubsectionspace
\subsection{Tracking and Sensing Analysis}  
\label{subsec:TrackingDataAnalysis}  
\reducesubsectionspace

To assess the accuracy of the estimated trajectories from both the VI-SLAM baseline system and the XR devices, we compare them against extrinsically calibrated ground truth trajectories. For this purpose, we utilize the EVO toolkit, a Python-based package designed for trajectory alignment and performance evaluation.
We report two key metrics in our analysis: Relative Pose Error (RPE) and Absolute Pose Error (APE). These metrics provide complementary insights into trajectory estimation performance.

\textbf{RPE:} 
The RPE evaluates the local consistency of the estimated trajectory by dividing it into fixed-length subtrajectories~\cite{zhang2018tutorial}. For each subtrajectory, the starting point is aligned with ground truth, and pose error is computed as the distance between the endpoint of the subtrajectory and its corresponding ground truth. This approach isolates drift errors within individual trajectory segments, ensuring that errors do not accumulate across subtrajectories. In our experiments, we set each subtrajectory length to 10 cm.

\textbf{APE: } 
The APE measures global trajectory alignment by computing the point-wise distance between the estimated trajectory and the ground truth at each timestamp~\cite{zhang2018tutorial}. Unlike RPE, APE aligns the entire trajectory once, meaning any accumulated errors directly affect overall accuracy. This metric reflects how well the estimated trajectory conforms to the ground truth over its entirety.

By reporting both RPE and APE, we capture different aspects of tracking performance: RPE highlights local consistency by isolating segment-level drift, while APE quantifies overall trajectory deviation. Together, these metrics provide a comprehensive evaluation of system accuracy and robustness.

\reducesectionspace

 \section{Experiment Setup}
\label{sec:ExperimentSetup}
\reducesectionspace
\begin{figure*}[th!]
\centering
\includegraphics[width=0.95\linewidth]{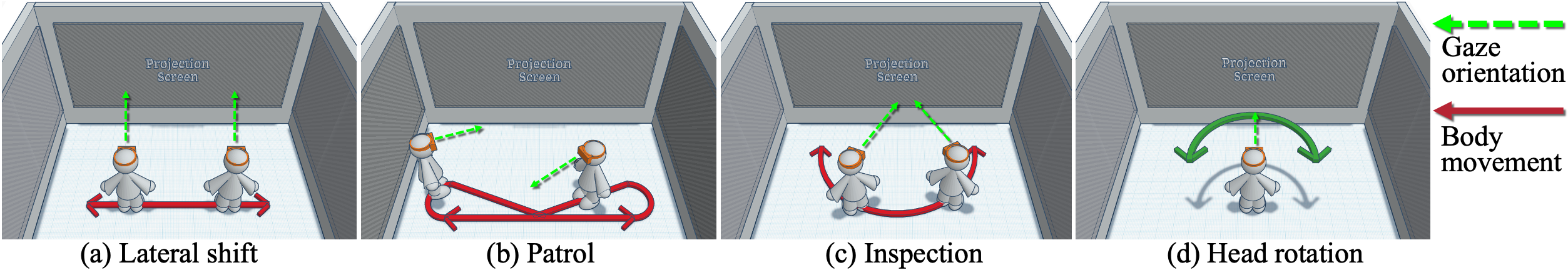}
\vspace{-0.5cm}
\caption{Movement patterns considered in our experiments: Lateral shift, patrol, inspect, and \change{head rotation}.}
\label{fig:MovementPattern}
\vspace{-0.3cm}
\end{figure*}

To systematically evaluate the tracking performance of XR devices, we conducted experiments using four distinct user motion patterns under controlled environmental conditions and movement speeds, as shown in~\cref{tab:experimentSetting}. These motion patterns were designed to represent typical XR usage scenarios while enabling precise quantitative comparisons. The experiments took place in a 6 m × 6 m mocap room equipped with 24 Vicon Vero v2.2 cameras and 4 Vicon Vantage v5 cameras. The Vicon system was calibrated to achieve an average positional error below 0.4 mm, ensuring highly accurate ground truth data for our measurements.

\vspace{-0.2cm}
\begin{table}[ht!]
\centering
\caption{\change{Experiment design with different motions and visuals.}}
\label{tab:experimentSetting}
\vspace{-5 px}
\begin{tabular}{|c|c|c|c|}
\hline
\textbf{Motion} & \textbf{Visual} & \textbf{Ex. ID} & \textbf{Note} \\ \hline
\multirow{2}{*}{\textbf{Rotate}} & Featureless & R-FL & \multirow{2}{*}
{\begin{tabular}[c]{@{}l@{}}$180^{\circ}$ in 4/4 time \\ at 50 and 75 BPM\end{tabular}} \\ \cline{2-3}
 & Feature-rich & R-FR & \\ \cline{2-3}
\hline
\multirow{2}{*}{\textbf{Shift}} & Featureless & S-FL & \multirow{2}{*}{\begin{tabular}[c]{@{}l@{}} side2side in 4/4 time \\ at 50 and 75 BPM\end{tabular}} \\ \cline{2-3}
 & Feature-rich & S-FR & \\ \cline{2-3}
\hline
\multirow{2}{*}{\textbf{Inspect}} & Featureless & I-FL & \multirow{2}{*}{\begin{tabular}[c]{@{}l@{}}semi-circle in 6/4 time \\ at 50 and 75 BPM\end{tabular}} \\ \cline{2-3}
 & Feature-rich & I-FR & \\ \cline{2-3}
\hline
\multirow{2}{*}{\textbf{Patrol}} & Featureless & P-FL & \multirow{2}{*}{\begin{tabular}[c]{@{}l@{}}side2side in 6/4 time \\ at 50 and 75 BPM\end{tabular}} \\ \cline{2-3}
 & Feature-rich & P-FR & \\ \cline{2-3}
\hline
\end{tabular}
\vspace{-0.4cm}
\end{table}

\reducesubsectionspace
\subsection{User Motion Patterns and Speeds}
\reducesubsectionspace

We implemented four distinct motion patterns (\cref{fig:MovementPattern}) to evaluate tracking robustness under varied kinematic conditions:

\textbf{Lateral shift:} Inspired by rhythm games and fitness applications, this pattern involves continuous side-to-side translation while maintaining a fixed forward gaze (\cref{fig:MovementPattern}.(a)). This motion produces gradual visual changes as the user's gaze direction remains stable during the lateral movement.

\textbf{Patrol:} 
Modeled after search-based interactions typical in first-person shooter games, this pattern requires users to traverse bidirectionally with abrupt U-turns, aligning their gaze with the direction of movement (\cref{fig:MovementPattern}(b)). The rapid changes in direction challenge the tracking system's stability during high angular velocity maneuvers and sudden shifts in visual flow.

\textbf{Inspection:} 
Drawing from AR/VR applications that necessitate moving with fixed focal points, such as object manipulation games like LEGO Bricktales and Angry Birds VR: Isle of Pigs, this pattern involves semi-circular walking while maintaining gaze fixation on a central target (\cref{fig:MovementPattern}(c)). This configuration provides consistent forward-facing visual input for head-mounted sensors while introducing lateral displacement.

\textbf{Head rotation:}  
\RV{R1C2}{
To simulate stationary experiences (e.g., 360° media viewing and meditation applications), users rotate their heads from side to side while keeping their bodies stationary (\cref{fig:MovementPattern}(d)). In our experiments, head rotation is restricted to the yaw axis, rather than pitch or roll, to ensure consistent device movement given that XR devices are vertically stacked in our testbed.}

\textbf{Speeds:} To standardize movement velocities across trials and evaluate performance at different speeds, we implemented metronome-paced movements. For patrol motions, movements alternated from side to side in 6/4 time at 50 and 75 beats per minute (BPM). Similarly, inspection movements followed a semi-circular path in 6/4 time at identical BPM settings. Lateral shift movements from side to side were executed in 4/4 time at 50 and 75 BPM, while head rotation involved complete $360^{\circ}$ turns in 4/4 time at the same BPM values. The 50 BPM setting approximated slow walking velocity, whereas the 75 BPM setting corresponded to jogging speed, providing distinct velocity conditions for comprehensive performance assessment.

\begin{figure*}[t!]
\centering
\includegraphics[width=0.95\linewidth]{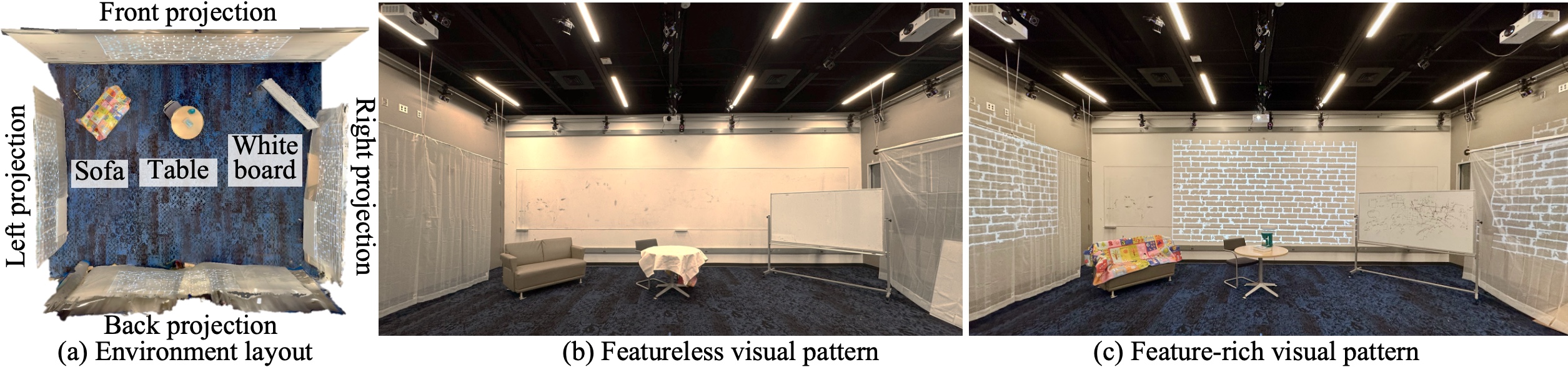}
\vspace{-0.4cm}
\caption{\change{Environment layout and visual patterns evaluated in experiments.}}
\vspace{-0.5cm}
\label{fig:VisualPattern}
\end{figure*}

\reducesubsectionspace
\subsection{Environment Setup and Visual Conditions}
\reducesubsectionspace
\RV{R4C4\&R3C4}{
The evaluation environment was established in a mocap space, where a study room was arranged with furniture including a sofa, table, chair, and whiteboard to introduce depth variation, as depicted in \cref{fig:VisualPattern}(a).
To assess the influence of environmental visuals on tracking performance, four high-brightness projectors (ViewSonic PA503HD, 4000 Lumens) were employed to alter the visual patterns on the surrounding walls.}

\change{
Two visual conditions were examined: featureless and feature-rich, as illustrated in \cref{fig:VisualPattern}(b,c). This selection was motivated by prior research in Structure from Motion and image localization, which demonstrated that environments lacking key points hinder pose estimation, whereas environments with evenly distributed key points enhance robustness~\cite{schonberger2016structure, ferranti2021can}.
}

\change{
\textbf{Featureless condition:} In the featureless setting (\cref{fig:VisualPattern}(b)), no visual content was projected onto the walls, resulting in an absence of salient features or key points. The sofa was left unchanged due to its solid color, while a white tablecloth was used to obscure patterns on the table. The whiteboard was oriented with the blank side facing the user. This configuration simulates an environment with minimal visual information available to tracking systems.
}

\change{
\textbf{Feature-rich condition:} In the feature-rich setting (\cref{fig:VisualPattern}(c)), a brick-wall pattern was projected onto the surrounding walls to provide dense and evenly distributed feature points for tracking. A small carpet with a block pattern was placed on the sofa, and the table was furnished with a book and coffee machine to increase visual features. The whiteboard was oriented with the side containing drawings facing the user.
}

\change{
By systematically varying these environmental visual conditions, we aim to analyze the impact of different levels of visual features on the accuracy and robustness of XR tracking systems.
}

\reducesubsectionspace
\subsection{Open-source SLAM Setup}
\reducesubsectionspace
\RV{R4C2\&R3C2, R3C5, R3C8, R1C3, R1C5}{
Due to restricted access to proprietary tracking modules and sensor data on commercial XR devices, we established an open-source tracking baseline using sensor data collected from the RealSense camera (\cref{subsec:TrackingSensingDataAcquisition}). This setup enables us to extract internal status, such as the number of keypoints extracted from camera frames, at runtime for correlation analysis with pose error.
While several prior works have developed and open-sourced tracking and mapping solutions for mobile AR~\cite{klein2007parallel,klein2009parallel,qin2018vins,li2024rd}, these systems are primarily designed for handheld devices with monocular cameras. This configuration fundamentally differs from XR headsets and glasses, which typically employ at least two tracking cameras. Accordingly, we adopted ORB-SLAM3~\cite{campos2021orb} (ORB3), an open-source VI-SLAM framework that supports stereo-inertial sensor configurations.
Because XR devices typically use VIO, a simplified form of SLAM that omits relocalization and loop closure~\cite{li2024rd}, we disabled these modules in ORB-SLAM3 to ensure a fair comparison.
} 

\reducesubsectionspace
\subsection{XR Device Selection and Specifications}
\reducesubsectionspace
In our experiments, we evaluated five XR devices with diverse tracking sensor configurations and processing units, as summarized in \cref{tab:XRdeivceSpecs}. \RV{R3C8}{The devices assessed were the HoloLens 2 (HL2), Magic Leap 2 (ML2), Meta Quest 3 (MQ3), Apple Vision Pro (AVP), XReal Air 2 Ultra (XR2U) and the open-source baseline ORB-SLAM3 on RealSense (ORB3)}. 
These devices were organized into two groups of three. Group 1 consisted of the MQ3, XR2U, and AVP, while Group 2 included the HL2, ML2, and AVP.
\RV{R4C3\&R3C3}{
The AVP was included in both groups to enable direct comparison with other leading commercial XR devices, given its anticipated superior tracking performance and to facilitate sensor data collection as mentioned in~\cref{sec:SensorDataCollection}.}

\RV{R4C3\&R3C3}{
For each group, the AVP was worn by the user, while the remaining two devices were mounted on 3D-printed heads. This protocol was adopted because the AVP is the only XR device employing Optic ID (an iris-based authentication system) and featuring downward-facing tracking cameras, which could be obstructed if mounted on a 3D-printed head.}
This configuration allowed each device to maintain identical environmental and kinematic conditions across all experiments while not interfering with each other's FoV, thereby ensuring fair and unbiased performance comparison.

\noindent In summary, our experimental setup involved evaluating five XR devices across two groups, subjected to four distinct motion patterns at two different velocities under \change{two} unique visual conditions. This configuration resulted in a total \change{32} different experiment configurations, during which tracking and sensing data were collected for subsequent evaluation and analysis. \textcolor{black}{The shared dataset\textsuperscript{\ref{GitHubRepo}} includes synchronized ground truth trajectories, device pose estimates, inertial and camera sensor data with timestamps, along with metadata on the experiment configurations, enabling thorough evaluation and facilitating subsequent research within the XR community.}
\reducesectionspace

\section{Experimental Results}
\label{sec:ExperimentResult}
\reducesectionspace

\begin{table}[t!]
\centering
\label{tab:OverallTrackingResult}
\caption{\change{Average tracking results of each device over all experiments and their pose errors w.r.t the Apple Vision Pro.}}
\vspace{-0.1cm}
\begin{tabular}{lllll}
\textbf{\begin{tabular}[c]{@{}l@{}}XR\\ Devices\end{tabular}} &
  \textbf{\begin{tabular}[c]{@{}l@{}}RPE\\ (cm)\end{tabular}} &
  \textbf{\begin{tabular}[c]{@{}l@{}}RPE+\\ w.r.t AVP\end{tabular}} &
  \textbf{\begin{tabular}[c]{@{}l@{}}APE\\ (cm)\end{tabular}} &
  \textbf{\begin{tabular}[c]{@{}l@{}}APE+\\ w.r.t AVP\end{tabular}} \\ \hline\hline
ORB3 & 1.57 & 304.8\% & 6.71 & 185.4\% \\ \hline
XR2U & 1.29 & 250.9\% & 8.44 & 233.1\% \\
HL2 & 1.43 & 278.1\% & 9.11 & 251.5\% \\
ML2 & 0.93 & 179.6\% & 6.11 & 168.8\% \\
MQ3 & 0.77 & 148.7\% & 4.52 & 124.8\% \\ \hline
AVP & 0.52 & 100\% & 3.62 & 100\% \\
\end{tabular}
\vspace{-0.6cm}
\end{table}

\begin{figure*}[th!]
\centering
\includegraphics[width=\linewidth]{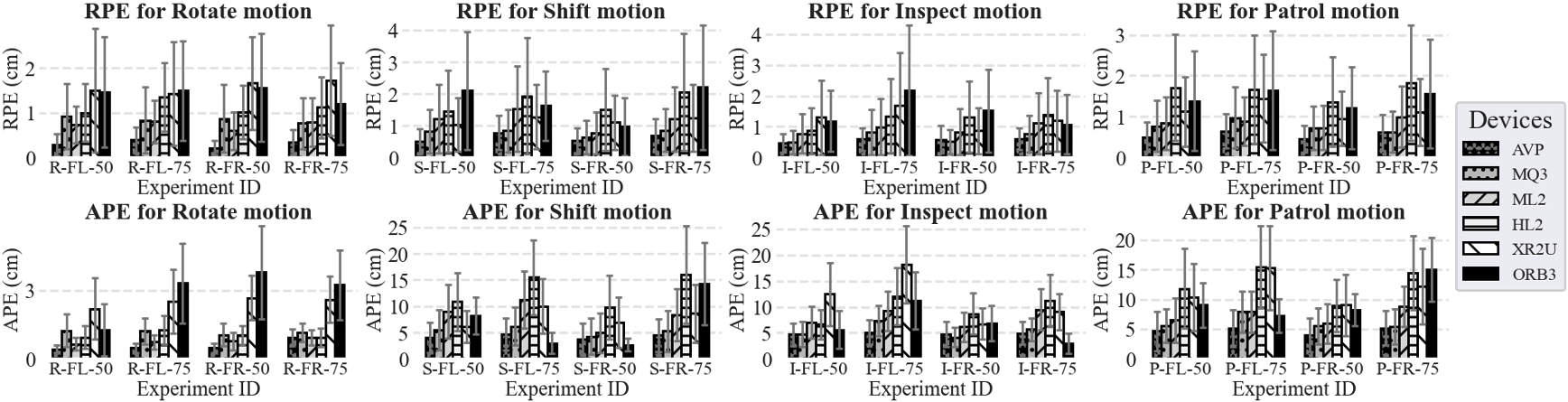}
\vspace{-0.8cm}
\caption{\change{Trajectory-level APE and RPE under different motion patterns, visual environments, and moving speeds.}}
\label{fig:BarChartofAllDevice}
\vspace{-0.2cm}
\end{figure*}

\begin{table*}[th!]
\caption{Specifications of the XR devices regarding their processing units, tracking sensors, and the OS version.}
\label{tab:XRdeivceSpecs}
\vspace{-0.1 cm}
\begin{minipage}{\textwidth} \centering
\resizebox{\textwidth}{!}{%
\begin{tabular}{|l|l|l|l|l|l|}
\hline
\textbf{Device}                                                        & \textbf{Year} & \textbf{Processor}                                                                                    & \textbf{Tracking Sensors}                                                                                          & \textbf{OS \& Version}                                                          & \textbf{Weight\footnote{Only contains the weight of the HMD and the head strap, excluding controllers or compute units}} \\ \hline
HoloLens 2                                                             & 2019          & \begin{tabular}[c]{@{}l@{}}Qualcomm Snapdragon 850,\\ Holographic processing unit (Gen2)\end{tabular} & \begin{tabular}[c]{@{}l@{}}2 stereo cameras\\ 2 periphery cameras \\ 1 IMU sensor\end{tabular}                     & \begin{tabular}[c]{@{}l@{}}Windows Holographic OS\\ 22621.1424\end{tabular}     & 566~g             \\ \hline
Magic Leap 2                                                           & 2022          & \begin{tabular}[c]{@{}l@{}}AMD quad-core Zen 2,\\ CVIP engine\end{tabular}                            & \begin{tabular}[c]{@{}l@{}}1 frontward camera\\ 2 periphery cameras\\ 2 IMU sensors (on the headsest)\end{tabular} & \begin{tabular}[c]{@{}l@{}}ML2 OS\\ v1.12.0 \\ B3E.241219.01\end{tabular}       & 260~g             \\ \hline
Meta Quest 3                                                           & 2023          & Qualcomm Snapdragon XR2                                                                               & \begin{tabular}[c]{@{}l@{}}2 stereo cameras\\ 2 periphery cameras\\ 1 IMU sensor\end{tabular}                      & \begin{tabular}[c]{@{}l@{}}Meta Horizon OS\\ v76.1021\end{tabular}              & 797~g\footnote{We replace the original head strap of Meta Quest 3 with BOBOVR M3 Pro for better stability}             
\\ \hline
Apple Vision Pro                                                       & 2024          & \begin{tabular}[c]{@{}l@{}}Apple silicon M2,\\ Apple R1\end{tabular}                                 & \begin{tabular}[c]{@{}l@{}}2 stereo camers\\ 2 periphery cameras\\ 2 downward cameras\\ 4 IMU sensors\end{tabular} & \begin{tabular}[c]{@{}l@{}}VisionOS 2.2\\ 22N5800a\end{tabular}                 & 671~g             \\ \hline
\begin{tabular}[c]{@{}l@{}}XReal Air 2 Ultra\\ + Beam Pro\end{tabular} & 2024          & Snapdragon 6 Gen 1                                                                                    & \begin{tabular}[c]{@{}l@{}}2 stereo cameras\\ 1 IMU sensor\end{tabular}                                            & \begin{tabular}[c]{@{}l@{}} Nebula OS\\ X4000\_X273\_2241129\_ROW\end{tabular} & 83~g              \\ \hline
\end{tabular}
}
\end{minipage}
\vspace{-0.6cm}
\end{table*}

% \Section{Expeirmental Results}
\cref{tab:OverallTrackingResult} summarizes overall tracking performance, while trajectory-level results are visualized in bar charts of \cref{fig:BarChartofAllDevice}. 
The correlation between pose error and sensor data is presented in \cref{fig:CorrelationAnalysis}.
Following the experimental setup, tracking performance was evaluated from three perspectives: device specifications, user movement patterns and speeds, and environmental conditions. 

\change{As shown in~\cref{tab:OverallTrackingResult}, across all experiments, the AVP consistently outperformed its peers, achieving an average RPE of 0.52~cm and APE of 3.62~cm. In contrast, other XR devices exhibited significantly higher errors. The HL2 showed the largest errors, with an RPE of 1.43~cm and an APE of 9.11~cm, approximately 2.8 times those of AVP, and demonstrated tracking performance comparable to handheld mobile phones manufactured in 2018-2019, such as Nokia 7.1 and Samsung Galaxy Note 10+ running AR Core, as observed in prior work \cite{scargill2022here}. The open-source baseline, ORB3, also performs badly, with a RPE three time that of the AVP. 
Meanwhile, the MQ3 demonstrated the best performance among non-AVP devices, but still exhibited RPE and APE values 48.7\% and 24.8\% higher than AVP. }

\RV{R2C4}{
These results highlight a substantial performance gap between the AVP and other commercial XR devices under challenging conditions. However, such differences may reflect varying design trade-offs: manufacturers often prioritize factors such as affordability, device weight, and battery life, which can influence sensor configurations and processing capabilities.
}

\reducesubsectionspace
\subsection{Impact of Device Specifications}
\reducesubsectionspace
As shown in \cref{tab:OverallTrackingResult} and \cref{fig:BarChartofAllDevice}, the AVP consistently outperformed all other devices across diverse motion patterns and environmental conditions. These performance differences are likely attributable to hardware specifications (\cref{tab:XRdeivceSpecs}).
A greater number of tracking cameras increases the total FoV, thereby reducing the risk of tracking errors or loss in environments with limited visual features. The AVP is equipped with six tracking cameras, compared to four in MQ3 and HL2, three in ML2, and only two in XR2U and RealSense D435i. Devices with fewer cameras, such as XR2U and ORB3, exhibited substantial performance degradation in featureless environments. 
\change{
For example, the XR2U's RPE increased from 1.20~cm in feature-rich environments to 1.69~cm in featureless environments during fast inspect movement (a 41\% increase).  The FoV of each camera also plays a critical role. While most XR devices use wide-angle tracking cameras, the ORB3 baseline equipped with RealSense cameras with a narrow FoV of $87^{\circ} \times 58^{\circ}$, resulting in approximately three times higher RPE compared to the AVP.}

Processing units also significantly impact performance. Spatial sensing and tracking require substantial computational power to process sensor data and estimate device pose with minimal motion-to-photon latency. Devices like the AVP and ML2 include dedicated co-processors designed specifically for sensor data processing. In contrast, older devices like the HL2, released in 2019, suffer from processing delays. For example, during the patrol motion with rapid head rotations and translation, the HL2’s estimated trajectory lagged behind ground truth, resulting in higher RPE and APE values. 
\change{
Specifically, HL2’s RPE during patrol in feature-rich environments at 75 BPM pace was 1.83~cm, 3 times higher than AVP’s 0.61~cm. Although the trajectory shape and scale are accurate, this lag causes local inconsistencies, making the HL2 perform poorly in head rotation experiments, with higher RPE and APE values compared to other devices.}

\reducesubsectionspace
\subsection{Impact of Motion Pattern and Speed}
\reducesubsectionspace
Our experiments demonstrate that both movement speed and movement pattern significantly influence tracking performance.

\textbf{Movement speed:} 
As shown in \cref{fig:BarChartofAllDevice}, faster movements generally caused increased tracking errors, both in RPE and APE, across all devices when the motion pattern and visual environment were held constant. 
\change{
For example, for the AVP in the shift motion with a feature-rich environment, APE increased from 3.71~cm to 4.50~cm, a 22\% increase. The effect was more pronounced in less capable devices: HL2’s APE under the same condition jumped from 9.80~cm to 16.06~cm, a 64\% increase.
This degradation in performance can be attributed to motion blur in camera frames caused by rapid movement, which reduces detectable keypoints available for tracking, thereby impairing overall accuracy.
}

\begin{figure*}[th!]
\centering
\includegraphics[width=0.95\linewidth]{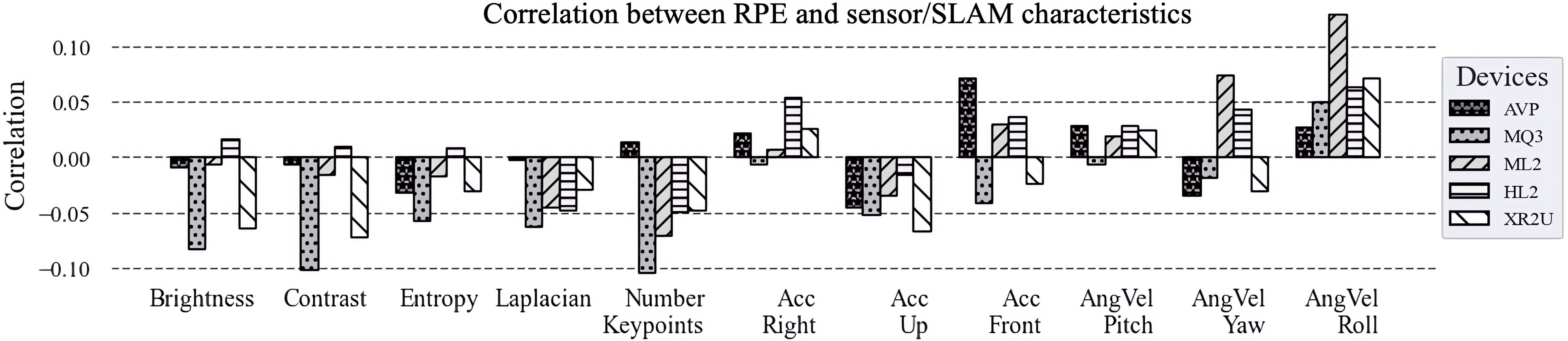}
\vspace{-0.4cm}
\caption{\change{Correlation analysis on SLAM runtime status and sensor readings at timestamp level.}}
\label{fig:CorrelationAnalysis}
\vspace{-0.65cm}
\end{figure*}

\textbf{Movement pattern:} 
The type of motion also influenced tracking performance. Head rotation motions generally exhibited lower APE but higher RPE. Head rotation motions exhibited relatively low APE but higher RPE across all visual environments and speeds. This is because, during head rotations, the user's body remains stationary, minimizing global inconsistencies. However, the rapid changes in orientation during rotation amplify local inconsistencies, which are reflected in the higher RPE values.

\change{
Lateral shift and inspection motions demonstrated consistent pose error ranges across most devices and conditions, except in featureless environments. This consistency is likely due to the gradual changes in visual content during these motions, allowing devices sufficient time to identify and track new keypoints, even at higher speeds.
Patrol motions yielded higher mean and variance in pose errors relative to inspection, likely because they involve more frequent and abrupt rotations and translations that rapidly alter the visual content captured by the device’s cameras.
}

\reducesubsectionspace
\subsection{Impact of Environmental Conditions}
\reducesubsectionspace
\change{
Our findings demonstrate that environmental feature density could significantly affect tracking performance.
\cref{fig:BarChartofAllDevice} shows that XR devices with fewer cameras were particularly sensitive to featureless environments. 
For instance, XR2U’s APE during inspection at high speed increased from 9.06~cm in feature-rich environments to 18.19~cm in featureless environments, a 101\% increase. 
By contrast, the AVP’s APE under the same conditions rose only slightly, from 4.98~cm in feature-rich environments to 5.15~cm in featureless environments (a 3\% increase), demonstrating its superior robustness.
Moreover, the negative impact of higher movement speed was exacerbated in the featureless environment. For example, HL2’s RPE during rotation movement in the featureless environment increased from 1.01~cm to 1.35~cm, a 34\% increase, while in the feature-rich environment, the APE increased from 1.01~cm to 1.12~cm, which corresponds to a 11\% increase.
}

\reducesubsectionspace
\subsection{Correlation Analysis}
\reducesubsectionspace
\change{
We analyzed the relationship between runtime-collected camera frame characteristics, VI-SLAM status, and IMU readings (from RealSense and ORB-SLAM3) and each device’s RPE by computing Pearson correlation coefficients. For IMU readings, which are inherently directional, we used the magnitude of each component to correlate with RPE. The results are summarized in \cref{fig:CorrelationAnalysis}.
}

\change{
\textbf{Image and SLAM characteristics: }
For each camera frame, we computed metrics including brightness, contrast, entropy, and Laplacian (the latter two quantify edge intensities), as well as the number of keypoints extracted by the open-source SLAM pipeline. Most XR devices exhibited negative correlations between these image characteristics and RPE, indicating that well-lit, feature-rich environments, where more significant keypoints can be extracted, enable better tracking performance.
}

\change{
\textbf{IMU characteristics:}
We examined acceleration components (Acc-Right, Acc-Up, Acc-Front) and angular velocities (AngVel-Pitch, AngVel-Yaw, AngVel-Roll), which reflect, respectively, bursts in translational and rotational user movement.
Most XR devices showed positive correlations between RPE and Acc-Right/Acc-Front, suggesting that pose error increases with lateral and forward acceleration. In contrast, Acc-Up displayed a negative correlation with RPE, potentially because vertical acceleration is synchronized with the user’s pace, and pose error tends to decrease during step transitions. Angular velocities in pitch, yaw, and roll generally correlated positively with RPE, indicating that rapid head rotations can increase pose error.
}

\change{
\textbf{Cross-device comparisons: }
Correlation patterns varied notably across devices. The AVP exhibited substantially weaker correlations with both image and IMU characteristics compared to other XR devices, likely due to its more advanced sensor configuration and more robust and consistent tracking performance. In contrast, the MQ3 demonstrated stronger correlations with image characteristics than with IMU readings, suggesting greater sensitivity to environmental visual conditions than to user kinematics.
}
\reducesectionspace

\section{Case Study}
\label{sec:CaseStudy}
\reducesectionspace
\RV{R1C4}{
This section presents a case study evaluating the feasibility of replacing mocap systems with AVP for pose error evaluation. We structure our analysis as follows:
Section~\ref{CaseStudy:Motivation} covers the motivation for using AVP instead of mocap, Section~\ref{CaseStudy:Implementation} details our implementation, and Section~\ref{CaseStudy:Results} analyzes our experimental results.
}

\begin{figure}[t!]
\centering
\includegraphics[width=0.72\linewidth]{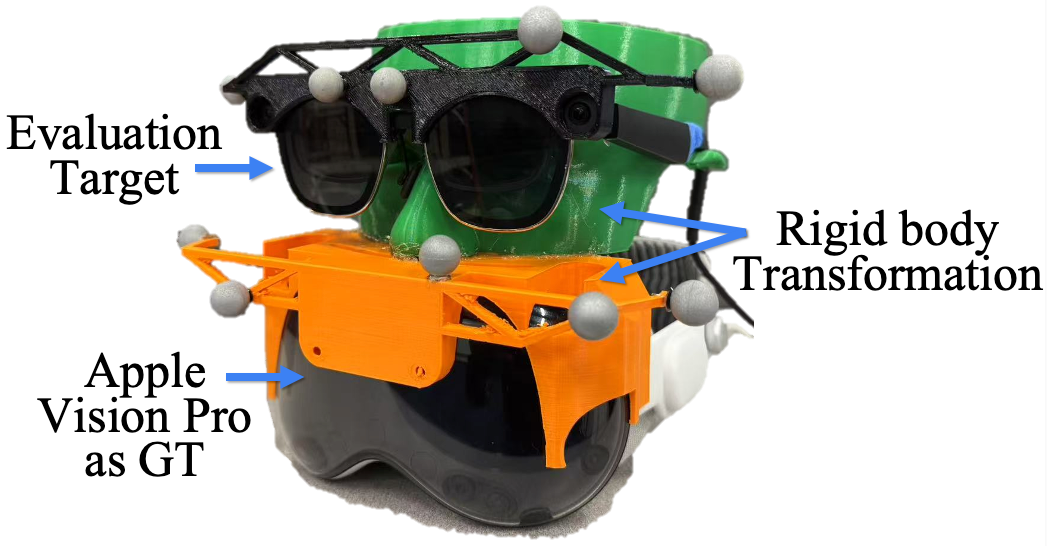}
\vspace{-0.52cm}
\caption{Using AVP as the ground truth to evaluate other XR device, where we place the target device on a fake head that is directly mounted on the AVP.}
\vspace{-0.42cm}
\label{fig:AVP-GT-Hardware}
\end{figure}

\reducesubsectionspace
\subsection{Motivation}
\label{CaseStudy:Motivation}
\reducesubsectionspace
\change{Traditional mocap systems, while accurate, require controlled environments with specialized hardware (e.g., infrared cameras, markers) and labor-intensive calibration. These constraints limit their practicality for real-world XR evaluations.}

\RV{R1C4}{
AVP presents a compelling alternative for ground truth trajectory estimation, offering two primary advantages. First, its self-contained inside-out tracking removes the need for external infrastructure, allowing flexible deployment across diverse environments. Second, our prior quantitative evaluation demonstrates that AVP achieves significantly lower RPE and APE than other consumer-grade XR devices, often by several factors. This combination of high tracking accuracy and operational flexibility positions AVP as a viable and practical ground truth source for trajectory validation.}

\reducesubsectionspace
\subsection{Apple Vision Pro Implementation as Ground Truth}
\label{CaseStudy:Implementation}
\reducesubsectionspace

\begin{figure}[t]
\centering
\includegraphics[width=0.92\linewidth]{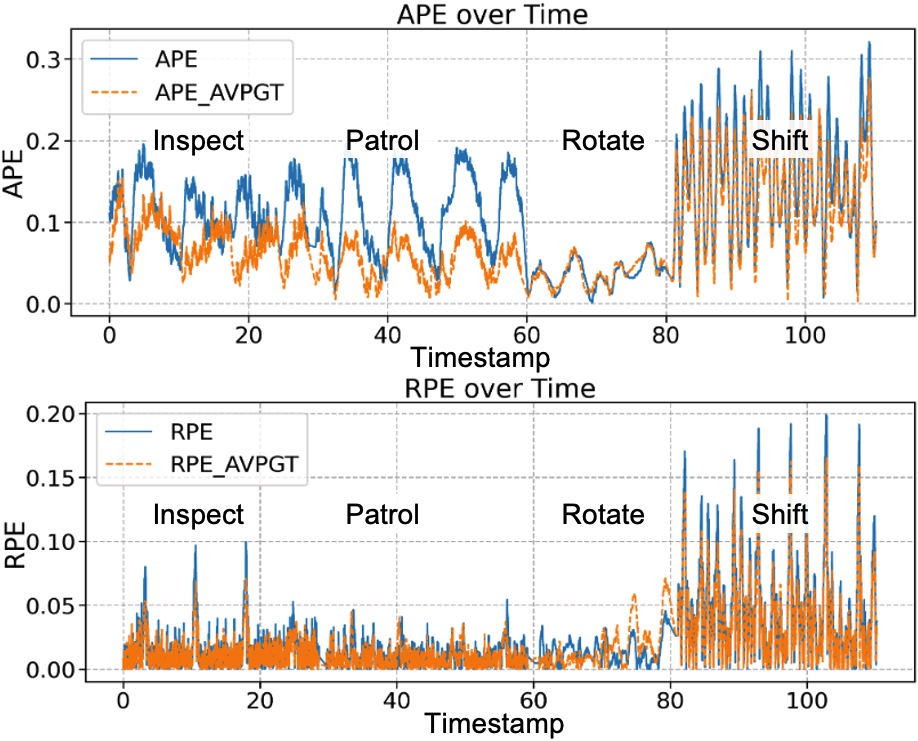}
\vspace{-0.5cm}
\caption{Comparison of the target device’s RPE and APE using either the mocap or AVP as the ground truth source.}
\vspace{-0.6cm}
\label{fig:AVPGT}
\end{figure}

\change{To use AVP as a ground truth source for other XR devices, it is essential to maintain a rigid and stable spatial relationship between the AVP and the evaluation target throughout the experiment. Therefore, we designed a XR device rig on the AVP as illustrated in \cref{fig:AVP-GT-Hardware}. In this configuration, the target device is mounted on a 3D-printed head that is directly affixed to the AVP, thereby ensuring a rigid transformation between the two devices.}

\change{
Consistent with our previous approach (\cref{fig:3D-print-heads-on-helmet}), we attach infrared markers to both the AVP and the target device. This allows us to simultaneously capture ground truth trajectories using the mocap system, providing a baseline for evaluating the feasibility of using AVP as the ground truth reference.}

\change{
For this case study, we select the XR2U as the evaluation target. We perform four user movement patterns illustrated in \cref{fig:MovementPattern} at a pace of 50 BPM within the feature-rich environment. Following the system calibration procedure described in Section~\ref{subsec:SystemSetupInit}, we calibrate the AVP and XR2U to compute the relative transformation between their device frames. This transformation is then applied to AVP’s trajectories to derive ground truth trajectories for the XR2U.}

\reducesubsectionspace
\subsection{Case Study Results}
\label{CaseStudy:Results}
\reducesubsectionspace
\change{
Figure~\ref{fig:AVPGT} compares the target device’s RPE and APE when evaluated using either the AVP or the mocap system as the ground truth source. The results yield $R^2$ scores of 0.830 for RPE and 0.387 for APE, indicating that while the AVP provides an accurate ground truth reference for measuring local pose error, it is less effective for measuring global pose error.
For APE, the AVP’s measurements are more consistent with the mocap system during rotation and lateral shift motions, but notable discrepancies arise in inspection and patrol movements, where the AVP tends to underestimate APE. We hypothesize that this underestimation occurs when both the AVP and the target device experience similar pose errors, thereby reducing the estimated APE.
}

\reducesectionspace

\section{Discussion and Implications}
\label{sec:DiscussionImplication}
\reducesectionspace

\RV{R2C2}{
Prior research~\cite{azuma1993tracking, bauer2007tracking, guan2023perceptual} has investigated tracking requirements for XR experiences and the impact of pose error on user perception, providing a foundation for interpreting our evaluation results.
For instance, Guan et al.~\cite{guan2023perceptual} quantified user tolerance for tracking errors, finding that AR applications have much stricter thresholds ($31.4 \text{ mm}^2 \pm 36.5 \text{ mm}^2$, or 3.2 mm drift) compared to VR ($7.6 \text{ cm}^2 \pm 7.0 \text{ cm}^2$, or 15.6 mm drift), primarily because AR overlays virtual content onto real-world references, making even small tracking errors highly perceptible~\cite{azuma1993tracking}.
}

\change{
\textbf{Our results indicate that, despite advances in XR technology, all five XR devices tested achieve average RPE within VR tolerance thresholds but exceed stricter requirements for AR applications}. This highlights a persistent gap in tracking precision for AR applications that demand high accuracy, such as AR-guided surgery or industrial maintenance~\cite{bauer2007tracking}.}
\change{
While it is well understood that minimizing abrupt user movements and providing feature-rich environments can reduce pose error, our findings underscore that developers cannot yet disregard these best practices. 
Also, the noticeably poorer performance observed in XR glasses suggests that, as more such devices enter the market, the burden on developers to carefully design applications, minimizing unnecessary user movement, will likely increase. Furthermore, developers must consider whether to implement applications in AR or VR, given users' heightened sensitivity to pose error in AR environments.
}

\reducesectionspace

\section{Limitations and Future Work}
\label{sec:FutureWork}
\reducesectionspace

\RV{R1C2, R3C6}{
\textbf{Device mounting and user motion:}
Our testbed evaluates multiple XR devices simultaneously by mounting them at different vertical heights (shown in \cref{fig:3D-print-heads-on-helmet}), which may subject each device to distinct motion trajectories, especially during rotational movements. To mitigate the impact of these differences, we limit head rotation to the yaw axis.
Also, the weight of the helmet and the 3D-printed heads is 1.4~kg. Together with three XR devices mounted and all markers and sensor module installed, the total weight added to the user's head could reach 3.1~kg, potentially hindering natural movement.  
In future work, we plan to design a more compact and lightweight mounting rig that minimizes vertical offsets and total weights while maintaining multi-device compatibility and increasing the user's comfort for a more natural movement.
Additionally, we will conduct more experiments with devices mounting at different positions to report mean and standard deviation of tracking performance across various heights.
}

\RV{R4C1\&R3C1}{
\textbf{Extrinsic calibration accuracy:} 
Our extrinsic calibration uses XR devices' output to compute the offset from the rigid body center to the device center. Although the calibration process is performed in a feature-rich environment with slow and steady movement, the estimated pose may still contain errors, potentially resulting in biased extrinsic calibration and affecting the overall tracking evaluation, especially for XR devices with poor tracking performance.
}

\change{
\textbf{System workload and resource contention:}  
Our current setup does not account for dynamic system load (e.g., CPU/GPU utilization) or rendering complexity, which can affect tracking performance under resource contention. 
Previous study~\cite{li2022timing} demonstrates that system resource contention significantly impacts program execution time and tracking accuracy. In future work, we will systematically vary workload and measure its effect on tracking accuracy.  This will allow us to quantify how rendering and processing demands influence tracking errors, leading to a more comprehensive understanding of device performance in realistic scenarios.
}

\change{
\textbf{Pose error metrics and user experience:}  
Although pose error metrics such as RPE and APE are standard for evaluating tracking performance~\cite{hu2024apple, zhang2018tutorial, grupp2017evo} , these metrics may not fully correlate with user-perceived quality~\cite{jiang2025remotevio, madsen2014wrong, huang2024replayar}. To address this gap, we plan to develop metrics that align more closely with human perception and user experience. By linking objective error measurements with human perception and application-specific criteria, we can ensure that our evaluation framework remains more relevant and actionable for designing future XR systems.
}

\reducesectionspace

\section{Conclusions}
\label{sec:Conclusions}
\reducesectionspace
\change{We introduced a novel testbed for benchmarking spatial tracking in XR devices, enabling the first systematic comparison of five SOTA XR devices under various environmental and kinematic conditions. Our results reveal substantial variability in tracking accuracy, with APE ranging from 3.62~cm to 9.11~cm and RPE spanning 0.52~cm to 1.43~cm across devices and environments. Correlation analysis highlighted the influence of environmental features and user motion on tracking performance, while our case study demonstrated the potential and limitations of using AVP as a ground truth reference. These findings advance the understanding of spatial tracking in XR and provide a foundation for developing more reliable and accurate XR systems.}

%\reducesectionspace

\acknowledgments{
\reducesectionspace
 This work was supported in part by NSF grants CSR-2312760, CNS-2112562, and IIS-2231975, NSF CAREER Award IIS-2046072, NSF NAIAD Award 2332744, a Cisco Research Award, a Meta Research Award, Defense Advanced Research Projects Agency Young Faculty Award HR0011-24-1-0001, and the Army Research Laboratory under Cooperative Agreement Number W911NF-23-2-0224. The views and conclusions contained in this document are those of the authors and should not be interpreted as representing the official policies, either expressed or implied, of the Defense Advanced Research Projects Agency, the Army Research Laboratory, or the U.S. Government. This paper has been approved for public release; distribution is unlimited. No official endorsement should be inferred. The U.S. Government is authorized to reproduce and distribute reprints for Government purposes notwithstanding any copyright notation herein.
 \reducesectionspace
 }

% %\bibliographystyle{abbrv}
%\bibliographystyle{unsrt}
%\bibliographystyle{abbrv-doi}
\bibliographystyle{IEEEtran}
% %\bibliographystyle{abbrv-doi-narrow}
% %\bibliographystyle{abbrv-doi-hyperref}
% %\bibliographystyle{abbrv-doi-hyperref-narrow}

\bibliography{references}

% \begin{appendices}
%     \include{10-Appendix}
% \end{appendices}

\end{document}